\documentstyle[12pt,aasms4]{article}
\def\etal{{\it et al.\/}}
\def\eg{{\it e.g.}}
\def\ltw{\>\hbox{\lower.25em\hbox{$\buildrel <\over\sim$}}\>}
\def\gtw{\>\hbox{\lower.25em\hbox{$\buildrel >\over\sim$}}\>}

\def \radm2{\rm rad/m$^2$}

\begin{document}

\vglue 0.1in

\title{\bf THE COMPLEX CORE OF ABELL 2199:} 

\title{\bf THE X-RAY and RADIO INTERACTION}

\author{\bf F. N OWEN}

\affil{National Radio Astronomy Observatory$^1$\\
P.O. Box O, Socorro NM 87801, USA}

\author{\bf J. A. EILEK}

\affil{Physics Department and Astrophysics Research Center \\
New Mexico Tech, Socorro NM 87801, USA}

\vskip 2.5in
$^1$The National Radio
Astronomy Observatory is operated by  Associated Universities, Inc., under
contract with the National  Science Foundation.







\begin{abstract}

The cluster Abell 2199 is one of the prototypical ``cooling flow''
clusters.  Its central cD galaxy is host to a steep-spectrum radio
source, of the type associated with cooling cores.  In this paper we
combine radio data with new ROSAT HRI data to show that conditions in
its inner core, $\ltw 50$ kpc, are complex and interesting.  Energy
and momentum flux from the radio jet have been significant in the
dynamics of the gas in the core.  In addition, the Faraday data
detects a dynamically important magnetic field there.  The core of the
X-ray luminous gas is not a simple, spherically symmetric cooling
inflow.  In addition, we believe the X-ray gas has had strong effects
on the radio source.  It seems to have disrupted the jet flow, which
has led to dynamical history very different from the usual radio
galaxy. This particular source is much younger than the galaxy, which
suggests the disruptive effects lead to an on-off duty cycle for such
sources. 


\end{abstract}

\section{Introduction}

Cluster cores are interesting places.  Many cluster centers have steep
X-ray profiles and low X-ray temperatures, which have lead to models
of spherically symmetric cooling flows (\eg\ the review of Fabian
1994).  It is becoming clear however that the central regions are more
complex than this simple picture.  There is evidence for transonic,
turbulent flows (\eg\ Baum 1992)
and for dynamically important magnetic fields (Taylor, Barton \& Ge
1994;  Eilek, Owen \& Wang 1997), at least in the inner $\sim 10-50$
kpc of strongly cooling cluster cores.  

In addition, when the central galaxies in cooling cores host radio
galaxies, these radio sources are not typical.  They are more diffuse in
appearance and have steeper radio spectra than most larger-scale radio
sources (\eg, Burns 1990; the halo of M87 is also a
nearby example of this class, \eg, Feigelsen \etal\ 1987).  It seems
very likely that these sources have a different dynamical history than the
larger-scale ones not generally found in cluster centers, and that
their history has been affected by their position in the center of a
dense cooling core. 

In this paper we combine X-ray and radio data to study the central
dynamics of one such cooling-core cluster, Abell 2199, with its
embedded radio galaxy, 3C338.  This cluster
seems quite normal dynamically, with no strong evidence of subclumps
in velocity data (Zabludoff \etal\ 1990).  It contains a central cD
galaxy, NGC 6166.  This galaxy appears to have a triple nucleus, which is
likely to arise from a projection of very eccentric orbits (Lauer
1986) rather than from a tightly bound central system.  The cluster
redshift is $z = .0312$ (Hoessel, Gunn \& Thuan 1980), so that 1 arcmin
corresponds to 36 kpc if $H_o = 75$ km/s-Mpc.  

The X-ray distribution from this cluster
is also quite regular, indeed rather dull, on large scales.  The
Einstein image (Forman \& Jones 1982) shows smooth, roughly circular
isophotes.  Jones \& Forman (1984) quote a core radius $\sim 210$ kpc
(converted to our ${H_o} = 75$ km/s-Mpc).  The smooth distribution shows
no particular signs of subclumping or merger activity.  This cluster
is one of the prototypical cooling cores:  the inner regions show the
characteristic X-ray excess relative to a 
King-model fit.  Arnaud (1985) finds a cooling radius $\sim 120$ kpc,
a central cooling time $\sim 0.8$ Gyr, and a mass inflow $\sim 90
M_{\sun}$/yr.  The more detailed cooling-flow model of Thomas, Fabian
\& Nulsen (1987) has similar numbers.  

NGC 6166 contains a radio source, 3C338.  It is small (radial extent
$\sim 35$ kpc)  moderate power ($8.8 
\times 10^{41}$ erg/s between 10 MHz and 100 GHz, data from Herbig \&
Readhead 1992, converted to $H_o = 75$ km/s-Mpc), steep spectrum
(the integrated 
spectrum is given in Herbig \& Readhead, and two-frequency spectral
index distribution in Burns, Schwendeman \& White 1983), and diffuse
in  appearance.  Thus it fits well in the class of cooling-core radio 
sources.  Newer radio data have been obtained by Feretti \etal\ (1993),
who present VLB images of a two-sided core jet, and by Ge \& Owen 
(1994), who made new VLA images with polarization and Faraday
rotation data. 

In this paper, we combine new ROSAT data with existing radio data to
explore the interesting dynamics of the core of A2199.  We find direct
evidence of the effect of the radio source on the central X-ray
luminous gas.  In this region, at least, we find that the X-ray gas
cannot be described by a simple, spherical cooling flow model.  We then
compare this radio source with standard models for ``normal'' radio
sources, and conclude that this source is relatively young and has
been disrupted by the surrounding gas.  We suspect that this scenario
may be typical of the class of cluster-center radio sources.

\section{The Data:  The Central Regions of A2199}

The intracluster medium (ICM) in Abell 2199 is smoothly distributed on
large scales, and 
appears to be sitting quietly in the potential well of the central
galaxy.  In Figure 1 we show an optical image of the central $\sim 4$
amin of the cluster (the data were taken by Owen \& White 1991).  The
elongation of the low surface brightness 
halo of NGC 6166 is apparent.  Lauer (1986) finds the position angle
of the optical isophotes $\sim 30^{\circ}$, for the inner $\sim 30$
asec; Owen \& White  find the same position angle for the 24.5
mag/asec$^2$ isophote, on a scale of $ \sim100$ asec $\sim 60$ kpc.
As we shall show, the large-scale X-rays are consistent with this. 
 
We obtained ROSAT HRI data of the inner $\sim 1000$ asec ($\sim 600$
kpc) of the cluster. A total of 47,522 seconds of useful integration
time was accumulated by the satellite in 1994 on Abell 2199. The observations
were made during two time periods, 3Feb94 to 5Feb94 and 31Aug94 to 6Sep94,
each with about half the total integration time. The $17\time 17$ amin
area we studied has 166,725 detected photons. Using 2 arcsec pixels the
peak on the image (at the nucleus of NGC6166) had 27 photons.
No identifications of
discrete sources with optical objects were found on the combined image
except for the core of the central galaxy. Thus no checks were possible
on the default coordinate system. However, the positional agreement between
the two parts of the integration are good, agreeing within about 2
arcsec in each coordinate for the brighter discrete sources in the
field, so
we have adopted the default coordinate system. The photon statistics
are not good enough to attempt to track any pointing wander as in
Morse (1994). However, the
discrete sources within 6 amin of the nucleus in the final image 
are consistent with FWHM of about
6 asec EW and slightly less NS. Since the peak on the
image is not very large, features much larger than this
are probably real.
At the resolution we have have used
to display the x-ray emission, all the features described are seen on
images made with either dataset.   

In Figure 2 we show 
a contour map of the field, smoothed to 56 asec resolution.  We note
that the isophotes remain smooth, with no particular evidence of
substructure, and show a only modest ellipticity.  We binned the
unsmoothed data
into circular rings to derive a luminosity profile, which is shown in
Figure 3.  The error bars in this figure assume Gaussian noise. From
$\sim 1$ amin to $\sim 10$ amin, the surface brightness follows a 
power law, $S_x(R) \propto R^{-1.2}$ if $R$ is the projected distance
from the center.   At smaller projected radii, our surface brightness
profile continues to rise inwards, but less rapidly.  The Einstein
data of Jones \& Forman (1984) show a 
steeper falloff on larger scales, $S_x(R) \sim R^{-2.2}$ for $R \gtw
5$ amin.  Thus,  a single power law is not a good fit to the entire
cluster.  We inverted the surface brightness with an Abel 
transform to find the radial emissivity profile of the cluster, out to
$\sim 10$ amin.\footnote{In
order to reduce edge effects in the transform, we artifically
extrapolated the measured $S_X(r)$, following the observed power law
well past 
the last measured data point.  This technique reduces the artificial
turnup at large $r$ brought about by edge effects, but of course adds
an unsupported assumption to the large-scale structure we derive for
scales $\gtw 200$ asec.}  
To get the gas density, we assumed a temperature  $T = 2 \times 10^7$K
(consistent with Thomas, Fabian \& Nulsen 1987) throughout 
the cluster.  (We note that the emissivity in 
the ROSAT band is not a sensitive function of the temperature, in
this temperature range; this should provide a reliable measurement
of the density).  We
estimated the error in the deprojection by adding noise to each data point,
chosen randomly in a $\pm  \sigma$ range around the data; we repeated
this random choice, and the Abel inversion, 100 times and determined
the mean and standard deviation of the result.  Our final density
profile is shown in Figure 4. The gas density derived from this analysis 
also follows a good power law, $n_X(r) \propto n^{-1.2}$ past $r \sim
60$ kpc.  Inside of this it rises more gradually, as $n_X(r) \propto
n^{-2/3}$, and approaches $n \sim
0.06$ cm$^{-3}$ as $r \to 0$.

In Figure 5 we show
the ellipticities, and position angles, of elliptical isophotes fitted
to the X-ray image using the IRAF program ELLIPSE. This program works
it way from the solution at one radius to the next by considering a 
radius larger (or smaller) by a given fraction. In this case the
fraction used was 0.1. For this reason the inner part of the profile
is oversampled and the errors are not independent. This does not
affect any of the later analysis. On scales
$\gtw 80$ asec ($\sim$ 50 kpc) the ellipticities and position angles
remain approximately steady, at $\epsilon \sim 0.2$ and $\theta \sim
35^{\circ}$, respectively.  
We note that the isophotes are not very
elliptical, so that our spherically symmetric Abel
inversion should be valid.  In addition, in Figure 6 we show an
overlay of optical contours on the X-ray image, which demonstrates that the
position angle of the large-scale X-ray isophotes
correlates well with the structure of the central galaxy. 

Thus, the outer regions of the X-ray gas seem to be sitting quietly in
the potential well of the galaxy.  The region inside $\sim 50-60$ kpc
is more interesting, however.    This is the region inside of which
the X-ray isophotes (Figure 5) change position angle and shape.
For $r \ltw 20$ asec the isophotes have $\theta \sim 80^{\circ}$; at
this point they  rotate
abruptly to $\theta \sim 15^{\circ}$; past here they gradually rotate
into alignment with the outer values.  The ellipticity also changes;
the very inner isophotes have $\epsilon \sim 0.2-0.3$; at $r \sim 20$
asec they become round ($\epsilon \sim 0$); past this they flatten slightly
to agree, again, with the outer values.  

In addition, the volume within $\sim 50$ asec
is that occupied by the radio source.  In Figure 7 we show the 5 GHz
image of 3C338 from Ge \& Owen (1994).  The radio core coincides with the
nucleus of NGC 6166 (Ge \& Owen 1994), and also with the peak of the
X-ray image using the coordinate system provided with the ROSAT database.  
A short jet, apparently ending in hot spots,  is
visible on either side of the core, extending $\sim 3$ kpc to a pair
of hot spots.  A two-sided nuclear jet has been detected on VLB scales
(Feretti \etal\ 1993).  The orientation of this jet coincides with the
inner jet and hot spots on the VLA image. The bright ridge, or
filament, south of the radio core 
does not coincide with any stellar feature; it is likely to be simply
a high-emissivity filament, as is common in other sources ({\it e.g.}
M87, Hines, Owen \& Eilek 1989; 3C442, Comins \& Owen 1991).  The  more
diffuse radio lobes extend to $\sim 35$ kpc from its core, in our
image and also in that of Burns \etal\ (1983).  Roland, Hanisch \&
Peltier (1990) searched but found no evidence for more extended, diffuse
emisison; the source seems to stop at $\sim 35$ kpc.  We  point
out that the diffuse lobes also show non-uniform, filamentary internal
structure. 

The radio data also shows that the inner core contains magnetized
thermal plasma (in addition to the magnetized relativistic plasma
within the radio source). 
The Faraday rotation data of Ge \& Owen (1994) show that rotation
measures $\sim 1000$ rad/m$^2$, ordered on scales $\sim 3$ kpc, exist
in the inner core.  In the next section we show that this is evidence
of a dynamically important magnetic field in the core.

The complex dynamics of the inner core is also apparent in the central
parts of the X-ray image.  
In Figures 8 and 9 we show X-ray emission from the inner core, at
higher resolution (7.5 arcsec), with the radio contours overlaid.
From both figures it is apparent that the radio source and X-ray
luminous gas know about each other.  Figure 8 shows that the inner
X-ray gas extends almost directly to the north, for $\sim 40$ asec, 
at a right angle to the radio jet direction. 
 Figure 9 shows that the very innermost X-ray gas is
elongated east-west about 20 asec, showing ``ears'' that coincide 
with the two radio
hot spots at the ends of the inner jet.  

Thus, the images suggest that the X-ray and radio sources are
interacting on scales $\ltw 50$ kpc in the inner core.  In the next
section we explore this interaction.

\section{ Discussion:  Dynamics of the Centre}

The data show that the central regions of A2199 are not simple.  The
radio source and the X-ray luminous gas are interacting strongly.
Existing ``standard'' models for each component (cooling flows for the
X-ray gas, jet-driven dynamics for the radio source) must be modified
to account for this interaction. 

\subsection{The Central Regions of the Cluster Plasma}

The dynamics of the central $\sim$ 50 kpc of A2199 are not well
described by a spherically symmetric, cooling-flow model.  We find
that the core contains a dynamically significant magnetic field, and
that the radio source has a strong effect on the dynamics of the gas
core. 

The polarization data of Ge \& Owen (1994) show that the core has a
significant, ordered RM distribution, and that  the RM  must come
from foreground gas.  In particular, the RM does not appear consistent
with a totally random magnetic field; as with other sources (\eg,
Taylor \etal\ 1994, or Eilek \etal 1997), the sign and magnitude of the
RM has a consistent, ordered pattern.  We identify the foreground gas
as the X-ray luminous gas in the cluster core.  Taylor \& Perley
(1993) and Ge \& Owen (1993) showed that
the rotation-producing gas in Hydra A and Abell 1975, respectively, is
neither within the 
radio source, nor from embedded emission-line clouds, nor in a mixing
layer between the radio source and the cluster gas.  Their arguments
also apply to Abell 2199; thus we take the X-ray bright cluster gas
to be the source of the rotation.

The data show that the RM has a typical magnitude $\sim 750$
rad/m$^2$; it has a typical order scale $\sim$ 3 kpc.  The depth of
the RM patch along the line of sight is likely to be comparable, also
$\sim 3$ kpc; as the source is larger than this, we are likely seeing
``patches'' or ``flux ropes'' of this scale, in front of the source.
If this typical RM comes from gas at density 0.02 cm$^{-3}$ (which we
take as a typical value for the region from $\sim 4$ to $ \sim 80$
asec), the mean 
line-of-sight magnetic field is $\langle B_{\parallel} \rangle \sim 15
\mu$G.  If we increase this by a $\sqrt{3}$ factor to account for
likely projection, we estimate the mean magnetic pressure in the lobe
region $\sim 2.8 \times 10^{-11}$ dyn/cm$^2$. 
For comparison, the pressure of
the X-ray luminous gas in this region is $p_g = n k T \simeq 5.6 \times
10^{-11}$ dyn/cm$^2$ (for typical values in this core, we estimate $n
\sim 0.02$ cm$^{-3}$ from our deprojection, and $T \sim 2 \times
10^7$K as above).   The typical magnetic pressure is thus significant
compared to the ambient gas pressure.  

The RM data also show one smaller, high-RM patch:  a region with RM
$\sim 1200$ rad/m$^2$, with scale $\sim 300$ pc, to the west of the
nucleus (which can be seen in Figure 3 of Ge \& Owen 1994).  If this
is also the line-of-sight scale, this filament has 
$\langle B_{\parallel} \rangle \sim 250 \mu$G, and a minimum $p_B \sim
2.5 \times 10^{-9}$ dyn/cm$^2$.  This is significantly higher than the
ambient X-ray gas pressure.  This suggests that the feature is
strongly overpressured, and either locally self-confined magnetically,
or else  transient.  It is also possible, of course, that the RM
of this filament is 
increased by local density fluctuations or a chance projection (with a
longer line of sight, if we are looking along a long filament), or
that the feature is somehow locally confined by its own field.  Again,
this says that the core plasma is magnetized, at a level which is
important to its dynamics.

We also find that the gas distribution of the inner core has been affected
by the radio source.  As we pointed out above, inside of $\sim$ 50 kpc the
distribution of the X-ray gas changes from the uniform ellipticity and
position angle it shows on larger scales.  In particular, the axes of
the radio jets and of the inner X-ray isophotes are clearly connected
(as in Figures 8 and 9).  These images suggest the following picture.

On the smallest scales, the radio jet is transferring momentum to the
X-ray gas, pushing it out and causing the ``ears'' of X-ray emission
around the jet.  For instance, Chernin \etal\  (1994) modelled
supersonic jets moving through a medium with a short cooling time.
They find that high Mach number jets can transfer significant momentum
to the ambient gas, {\it at the head of the jet}.  Their calculation
may be relevant here.  Picking $n = 0.05$ cm$^{-3}$ for the inner few
kpc, $T = 2 \times 10^7$K and using the Raymond \etal\ (1976)
emissivities gives a cooling time $\sim 130$ Myr.  This is only a few
times larger that the dynamical age of the source, which we argue in
the next section is on the order of tens of Myr.

In addition, The X-ray isophotes are elongated to the immediate north
of the jet, opposite to the direction of the radio lobes.  This
suggests that power from the radio source has been deposited in the
inner X-ray gas, causing this structure.  
Quantitative estimates of the energetics of the core seem consistent
with this.  Recall $P_{rad} \sim 8.8 \times 10^{41}$ erg/s.  The beam
power is most likely much larger (\eg, Eilek \& Shore 1989).  If we
guess $\sim 1$\% as a typical  efficiency, the beam power is  $P_b
\sim 10^{44}$ erg/s.  Much of this beam power 
will be deposited in the ambient gas, as well as in the radio lobes
themselves.  We first note that the bolometric luminosity from the
X-ray gas, using the Raymond, Cox \& Smith (1976) emissivities, is
$L_{core} \sim 2.4 \times 10^{43}$ erg/s.  (We calculated this for the
inner 35 kpc of the X-ray core, the size of the radio lobes.)
This is very similar to the  
likely $P_b$ value.  Thus the beam power clearly has a strong effect
on the local thermal balance of the X-ray gas.

We can also estimate the energy 
content of the inner 35 kpc of gas.  Its thermal energy content is
$U_x = { 3 \over 2} p_x \simeq 2.6 \times 10^{59}$ erg; its
gravitational potential energy will be similar.  The radio beam
deposits this much energy in a time  $U_x / P_b \simeq 81 / P_{44}$
Myr.  This again argues that the radio source has had a significant
effect on the energetics of the inner core, and can have heated the
``northern extension'' of the X-ray gas enough to move it upwards in
the local gravitational potential.

\subsection{The Cluster-Center Radio Source}

The radio source 3C338 is typical of steep-spectrum cluster-center
radio sources.  We suggest that it has developed from a jet which has
been severely disrupted, by the conditions in the central cooling core
of this cluster.  It follows that this is a relatively young source;
and that such cluster-center galaxies must have an ``on/off'' cycle of
radio activity.

We begin our argument with the appearance of the source.  It is
more diffuse than standard Type I or Type II radio galaxies; it shows
neither strong external hot spots nor a directed tail flow emanating
from the core.  It does contain an active VLB jet, which connects to the
3-kpc scale VLA inner jets but does not continue into the
lobes.  Its diffuse appearance suggests that it has grown more as a
``bubble'', driven by its internal energy, than as a standard source
driven by directed jet flow.  The lobes are not particularly uniform.
They are inhomogeneous, breaking up into filaments as in Figure 7;
there is one bright ridge to the south of the core; and they are
located to one side of the central core (opposite to the
X-ray extension).  

We also point out that the radio plasma must be separate from the
X-ray luminous gas.  If the two plasmas were well mixed, the high
rotation measure found by Ge \& Owen (1994) would depolarize the
source (for example, a rotation measure of 750 rad/m$^2$ gives a
Faraday depth $\sim 1.8$ radians at 6 cm; this would easily depolarize
the source).  Thus the radio plasma cannot be well-mixed with the
local X-ray gas.  This argues against diffusion as the origin of this
source, and supports our suggestion of a `separate `bubble''. 

The pressure within the radio
source is consistent with our picture.  We calculated the minimum
pressure for the bright filament to the south of the radio
core, and also for the diffuse lobes. In doing the calculation, we
took the high-frequency spectral index $\alpha = 1.7$ from Burns
\etal\ (1983), but assumed the spectrum flattened to $\alpha = 1.0$ between
$10^7$ Hz and $10^9$ Hz.  (This give a more conservative estimate for
the minimum pressure.)  We also assumed a uniformly filled source, and
equal energy densities in relativistic protons and electrons.  We
found $p_{min} \simeq 1.2 \times 10^{-10}$ dyn/cm$^2$ for the bright
filament (corresponding to $B_{min~p} \simeq 42 \mu$G), and $p_{min}
\simeq 4.8 \times 10^{-12}$ dyn/cm$^2$ for the 
diffuse regions (giving $B_{min~p} \simeq 8 \mu$G).  Thus, the bright
filament is overpressure relative to the X-ray background (recall $p_x
\sim 5.6 \times 10^{-11}$ dyn/cm$^2$ is typical of the inner $\sim$ 40
kpc), while the lobes have $p_{min} < p_x$.  The latter condition is
compatible with pressure balance, because the true pressure in the
radio source can exceed $p_{min}$ easily; this will occur either if
the source is inhomogeneous (as it clearly is), or if it is not
exactly at the minimum pressure condition (which is physically quite
possible).  We suspect that overall pressure balance is maintained,
and that the bright filament to the south is a transient feature,
currently overpressured, for instance due to strong turbulence in the
region. 

What, then, is the dynamical history of this source?  We suspect that
the nuclear jet became unstable and disrupted severly, at
approximately its current position of $\sim 3$ kpc from the core.  We
are not aware of numerical simulations which specifically address this
situation.  Hardee \etal\ (1992) modelled jets propagating in 
atmospheric gradients.  Loken \etal\ (1994) modelled jets propagating
in a strong cooling inflow.  Both found that jets with lower  Mach
number  can be stalled out or suffer strong instabilities which
dramatically slow their propagation.  After the instabilitiy has
developed, the material flowing through the jet fills a ``lobe'' or
``bubble'' which grows only slowly thereafter. This bubble has now
reached the lobe size, $\sim 35$ kpc.  While neither of these
calculations addresses the conditions which we find in A2199 -- a
magnetized, probably turbulent, ambient medium -- we suspect similar
evolution may occur in this case.

If we adopt this model we can estimate the dynamical age of the radio
source.  One clue comes from the 3 kpc length of the jet.  Following
Scheuer (1974), we apply momentum flux balance at the end of the jet.
This predicts that a jet of opening angle $\Omega$, beam power $P_b$
and speed $v_b$ propagates into an ambient medium of density $\rho_x$
at a rate given by
$$
D(t) \simeq \sqrt{2} \left( { P_b \over \Omega v_b \rho_x}
\right)^{1/2}
\eqno(1)
$$
where  $\Omega v_b$ relates the beam power to its 
momentum flux.  Models of Type II RS (Eilek 1997a) suggest $\xi = 100
\Omega v_b/ c \sim 1$, and we use this scaling here as well.  Taking
$n_x \simeq 0.02$ from our X-ray data, the jet length expression
becomes 
$$
D(t) \simeq 0.84 \left( { P_{44} \over \xi} \right)^{1/4} t_{Myr}^{1/2}
~~ {\rm kpc}
\eqno(2)
$$
for propagation into a uniform medium at the density of the central
X-ray gas. 
This predicts that the jet reaches 3 kpc in $t_{jet} \sim 17$ Myr if
$P_{44} / \xi = 1$.  We suspect that it reached this point in this
time, then suffered strong instabilities and disruption; thereafter
mass and energy transport continued, but not as a collimated flow.
The source must therefore be at least as old as this.

A second constraint on the age comes from the expansion of the lobes.
We envision them growing due to the internal pressure of the plasma
which has passed through the end of the (now stalled) jet.  Following
Eilek \& Shore (1989), we describe the evolution of such a lobe which
expands at approximate pressure balance with its surroundings:
$$
V(t) \simeq {P_b \over p_x} t
\eqno(3)
$$
Taking the X-ray pressure, $p_x \simeq 5.6 \times 10^{-11}$
dyn/cm$^2$, with $V = 4 \pi R^3 / 3$ (ignoring non-sphericity), 
gives an estimate of the linear scale of the lobe:
$$
R(t) \simeq 7.8 \left( P_{44} t_{Myr} \right)^{1/3} ~~ {\rm kpc}
\eqno(4)
$$
so that the lobes have reached their current 35 kpc size in 
$t_{vol} \sim 90$ Myr if $P \simeq 10^{44}$ erg/s.  Our picture is
self-consistent, in that this time is longer than the
time needed for the central jet to reach its current scale.  

One caveat here is that these calculations assume propagation into a
uniform external medium.  In reality the ambient X-ray gas has a strong
density and pressure gradient, as is apparent from Figures 3 and 4.
This will accelerate both the jet propagation and the lobe growth.  We
are not aware of any specific models of such propagation, but expect
such affects to, say, reduce $t_{jet}$ and $t_{vol}$ by a factor of 2
or so. 
In addition one might expect bouyancy to be important in the lobe
growth.  This is also not included; we can estimate its effect by
arguing that bouyant velocities will be no larger than the local
gravitational speed.  Fisher, Illingworth \& Franx (1995) measure the
velocity dispersion $\sigma \sim 270$ km/s for NGC 6166; from this we
can estimate $t_{bouy} \sim 35$ kpc $/ \sqrt{3} \sigma \sim 60$ Myr.
This is comparable to $t_{vol}$, suggesting that bouyancy will also reduce
the growth time by an order-unity factor.  

Another possible estimate of the source age is from spectral
steepening.  Burns \etal\ (1983) take 400 MHz as the turnover
frequency.  If this applies throughout the source, we derive
synchrotron ages of $\sim 70$ Myr for the diffuse lobes (with
$B_{min~p} \sim 8 \mu$G), and $\sim 6$ Myr for the bright filament
(with $B_{min~p} \sim 42 \mu$G).  The larger of these numbers appears
consistent with our dynamical ages.  We caution, however, that
spectral steepening by itself is not a reliable estimate of source
ages; Eilek (1997b) demonstrates this for Type I sources, and Eilek,
Melrose \& Walker (1997) discuss an alternative interpretation of the
spectrum.  Thus, we regard the coincidence of this estimate with our
dynamical ages as interesting, but not definitive.

Thus, our dynamical picture suggests that the source is young; it has
taken  no more than several tens of Myr to reach its current size. 
This is significantly younger than the age of the parent system.

\section{Conclusions}

In this paper we presented new X-ray data which shows evidence of
a complex interaction between the two plasmas (relativistic, radio
bright, and thermal, X-ray bright) in the core of A2199.  We draw two
major conclusions from this.

First, the core of this prototypical ``cooling flow'' is a complex place.
The magnetic energy density (and one suspects turbulent energy
densities) are comparable to that of the thermal gas.  The radio
source is an important source of energy to the X-ray gas in the core.
Thus, this is not a simple symmetric cooling flow, at least on these
scales.  Second, the radio source shows signs of being disrupted by
the ambient gas.  It remains unmixed with the X-ray gas, but has a
different dynamical history than most radio galaxies.  We suspect this
is a clue to the unusual nature of steep-spectrum, cluster-center
radio sources.

It is worth noting that M87 is another very similar example of this
phenomenon.  The large-scale radio halo of M87 (\eg, Feigelsen \etal\
1987) has a similar, diffuse appearance, and similar linear scale.
The inner radio source of M87 has a jet which clearly disrupts on a
scale of a few kpc (\eg, Owen, Hardee \& Cornwell 1989).  The total
radio power of M87 is very close to that of 3C338 (Herbig \& Readhead
1992), although the minimum pressure in the diffuse radio lobes is
lower for M87 (Feigelsen \etal\ 1987). The inner regions of the X-ray
halo of M87 have a similar run 
of pressure (Nulsen \& B\"ohringer 1995), and also show evidence of
dynamical interaction with the radio galaxy (B\"ohringer \etal\ 1995).
Finally, the nuclear gas -- at least on the few kpc scale -- is also
strongly magnetized (Owen, Eilek \& Keel 1990).  Thus, at least one
other system seems very similar to the one we have studied here

Finally, we note that steep spectrum radio sources are common in cD's
in cooling cores;  Burns (1990) detected such sources in $\sim 2/3$ of
his sample.  While most have not been studied in detail, we suspect
that 3C338 and M87 are typical members of this class.  If this is the
case, the central cD must be radio-active for most of its life.  This
can only be reconciled with the relatively young age we deduce for
3C338 ($\sim 30 - 100$ Myr) if the parent galaxy has frequent,
short-lived radio-active periods.  Perhaps the instabilities which we
see now disrupting the jet, eventually shut it off totally, and after
a quiescent period the system restarts itself?

\bigskip
We thank Chris Loken for useful discussions on jet disruption, and
Fang Zhou for his help with the data.  JE was partially supported by
NASA grant NAG51848 and NSF grant AST-9117029.

\clearpage

\centerline{\bf Figure Captions}

\figcaption[] { Optical image,  obtained by Owen \& White (1966), of
the central region of the  cluster Abell 2199.   The cD galaxy NGC
6166 dominates the central region of the cluster.}

\figcaption[]  {The ROSAT HRI image of Abell 2199, smoothed to 56 asec
resolution. This choice gives a Gaussian beam of 1 amin$^2$.  The
units in this figure are counts per beam. Note the lack of
substructure, and the constancy of the 
isophote position angles on these scales.}

\figcaption[]  { The X-ray surface brightness of Abell 2199, obtained
by binning the unsmoothed data into circular rings.  The units of this
figure are counts per pixel, where each image pixel is 2 asec square.
Note the shoulder at $r \sim 80$ asec, which is approximately the
extent of the central radio source 3C 338.}

\figcaption[]  {The X-ray density of Abell 2199.  We derived this
using an Abel transform on the data in Figure 3, and assuming a
uniform temperature $T \simeq 2 \times 10^7$K. Note the shoulder in
the density distribution at $\sim  80$ asec.}

\figcaption[]  {(a) The ellipticities of the X-ray isophotes.  In this
figure and the next, the profile is oversampled close to the nucleus,
and therefore the errors in that region are not statistically
independent. Note 
the increasing elongation from $r \sim 0$ to $r \sim 16$ asec, the
sudden drop there to nearly round isophotes, and the connection past
that point to the larage-scale $e \sim 0.2$ value.  (b)  The position
angle of the isophotes.  Note the dramatic difference between the
inner isophotes, whose position angle is at right angles to the jet of
the radio source, and the outer isophotes, whose position angles agree
with the orientation of the cD galaxy.}

\figcaption[]  {An overlay of the optical image (contours) and the
X-ray image (color), demonstrating the good agreement of the
orientation of the stellar galaxy and the X-ray bright gas on these
scales.} 

\figcaption[]  {The 20 cm  radio image of 3C338, from Ge \& Owen
1994. The radio core is in the center of the picture, and is the
source of short jets which terminate in the hot spots at $\pm 3$ kpc
to either side of the core.  The grey-scale display is
proportional to the tenth root of the intensity, in order to emphasize
the fine structure and diffuse emission more clearly than is shown in
Ge \& Owen.  The core coincides with the galactic
nucleus and the VLB core and jet seen by Feretti \etal (1993).  The
bright filament to the south is not a jet; it seems to be simply a
high-pressure shock or flux rope in the radio plasma.}

\figcaption[]  {An overlay of the radio image (contours) and the X-ray
image (color).  This figure illustrates the $r \ltw 35$ kpc scales, on
which the X-ray gas is elongated north-south and seems to be strongly
affected by the radio source.}

\figcaption[]  {An overlay of the radio and X-ray images, similar to
Figure 8, but illustrating the smallest scales ($\ltw 5 $ kpc, seen as
the orange-red central region), on which the X-ray gas
is elongated east-west, and seems to be receiving momentum transferred 
from the radio jet.}

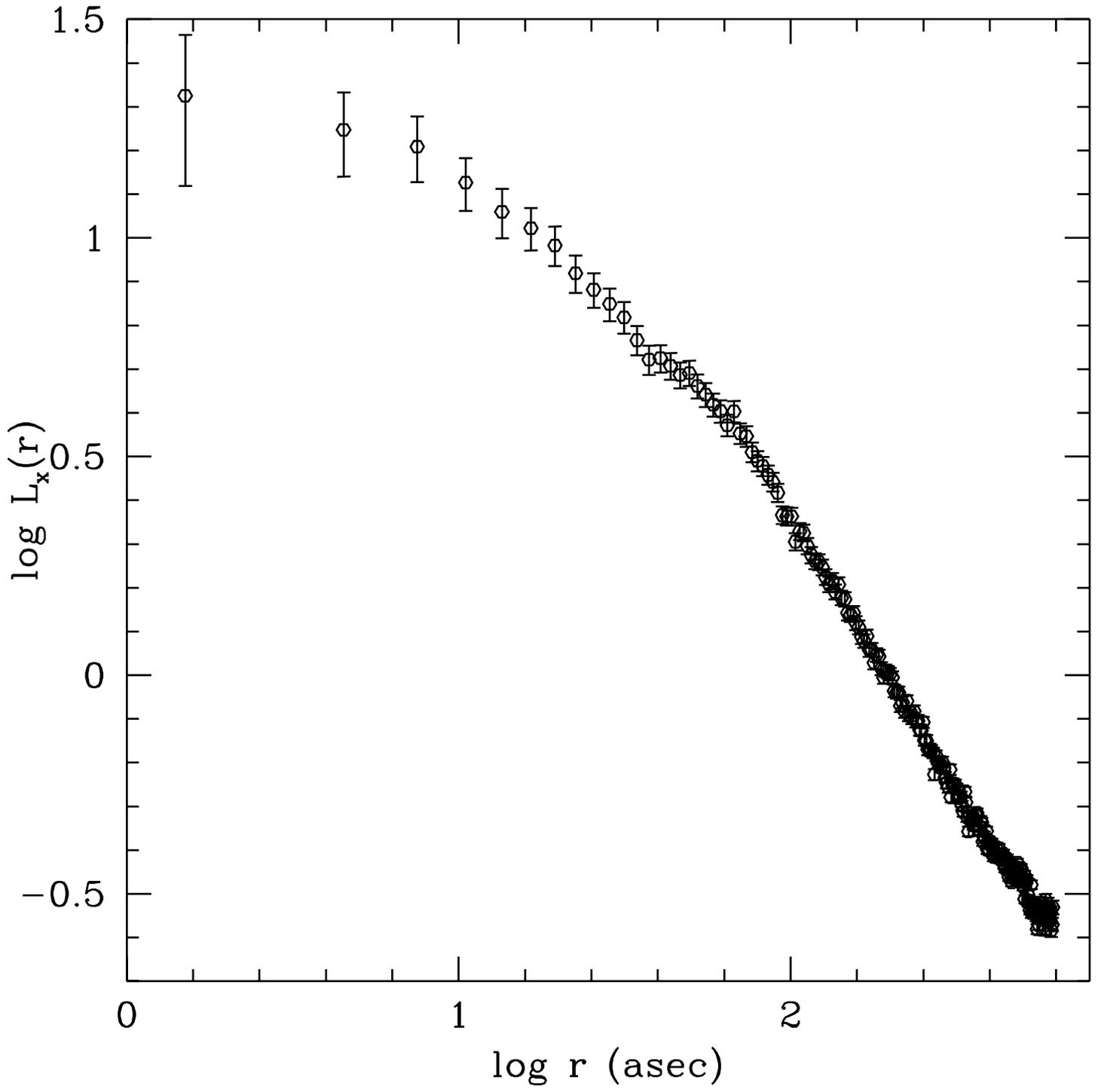
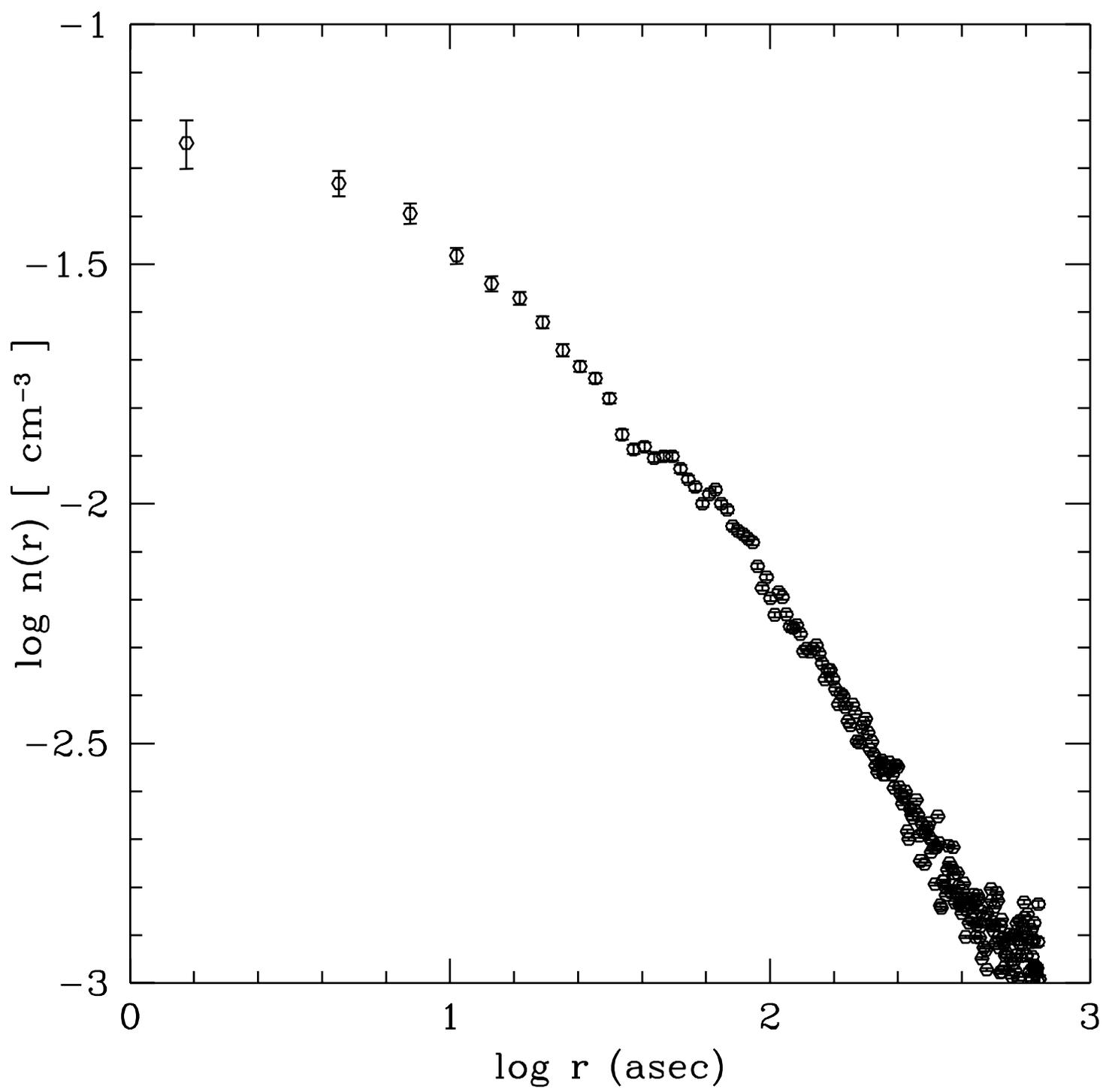
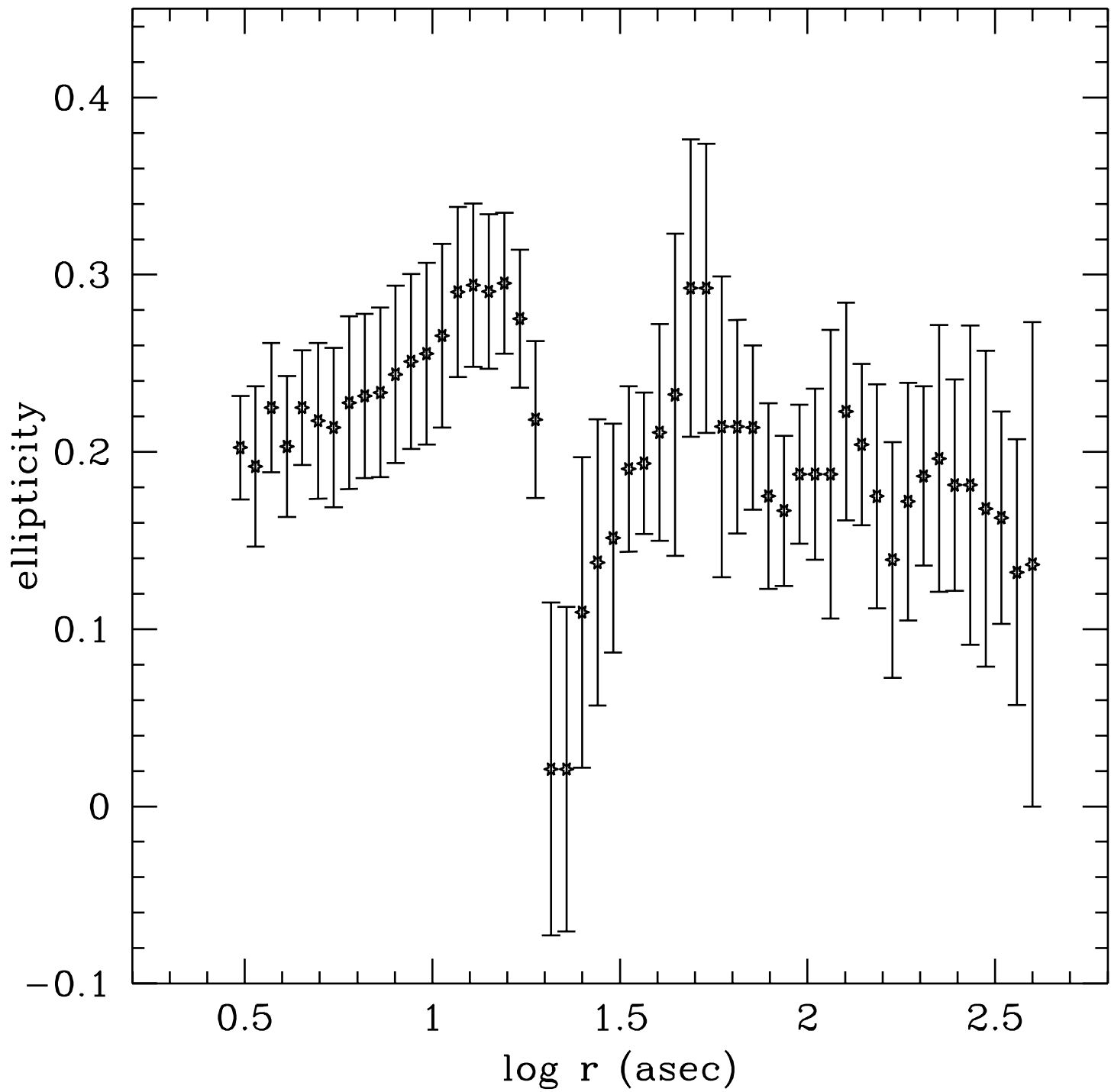
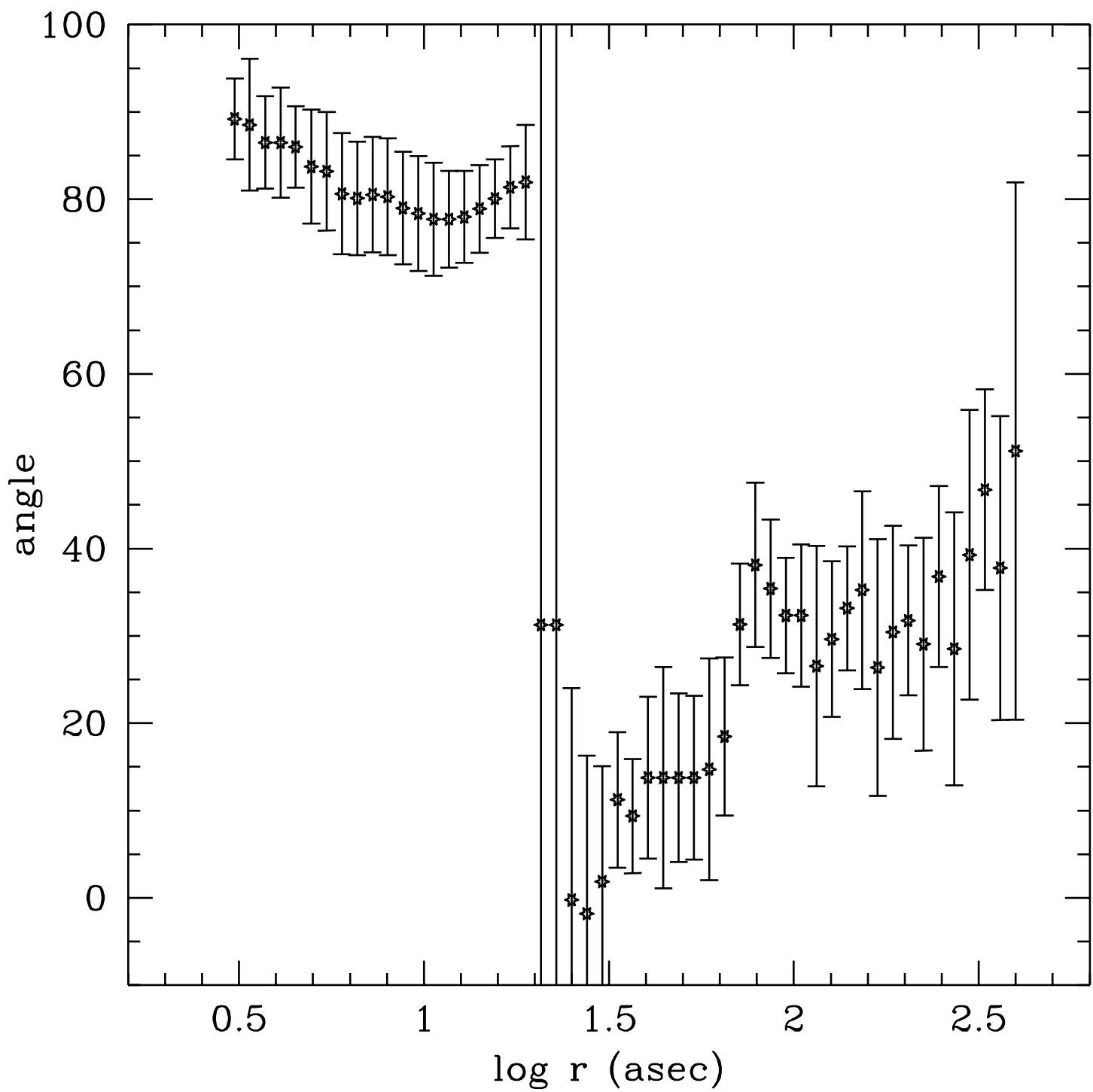
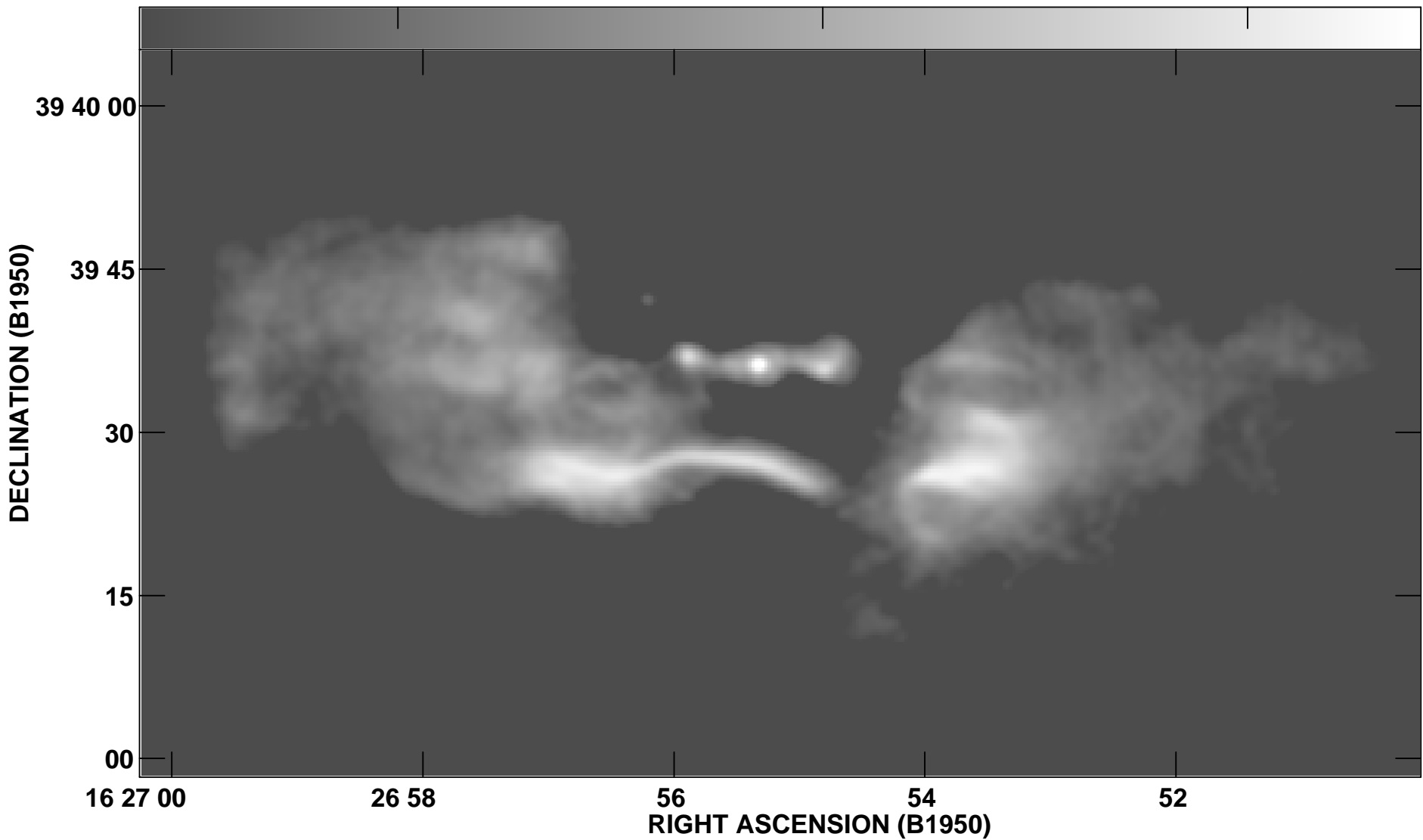
\end{document}